# Intrinsically Ultrastrong Plasmon-Exciton Interactions in Crystallized Films of Carbon Nanotubes

Po-Hsun Ho, Damon B. Farmer, George S. Tulevski, Shu-Jen Han, Douglas M. Bishop, Lynne M. Gignac, Jim Bucchignano, Phaedon Avouris[1], Abram L. Falk[2]

*Dept. of Physical Sciences, IBM T. J. Watson Research Center, Yorktown Heights, NY 10598*

**In cavity quantum electrodynamics, optical emitters that are strongly coupled to cavities give rise to polaritons with characteristics of both the emitters and the cavity excitations. We show that carbon nanotubes can be crystallized into chip-scale, two-dimensionally ordered films and that this new material enables intrinsically ultrastrong emitter-cavity interactions: rather than interacting with external cavities, nanotube excitons couple to the near-infrared plasmon resonances of the nanotubes themselves. Our polycrystalline nanotube films have a hexagonal crystal structure, ~25 nm domains, and a 1.74 nm lattice constant. With this extremely high nanotube density and nearly ideal plasmon-exciton spatial overlap, plasmon-exciton coupling strengths reach 0.5 eV, which is 75% of the bare exciton energy and a near record for room-temperature ultrastrong coupling. Crystallized nanotube films represent a milestone in nanomaterials assembly and provide a compelling foundation for high-ampacity conductors, low-power optical switches, and tunable optical antennas.**

When optical emitters are strongly coupled to a cavity, they hybridize with it via rapid energy exchanges known as vacuum Rabi oscillations (1). If the Rabi-oscillation frequency ($\Omega/2$) is so fast that it approaches the resonance frequencies of the emitters ($\omega_0$) and the cavity, the system has then reached ultrastrong coupling (2). Instead of the cavity and emitters exchanging energy one quantum at a time, a single cavity excitation can then borrow energy from the vacuum field and excite two or more emitters. Ultrastrong coupling could lead to single-photon

---

nonlinearities that provide a pathway from fundamental concepts in quantum electrodynamics to advanced telecommunications hardware (3, 4).

There are many emitter-microcavity systems that have achieved strong or ultrastrong coupling. The emitters can range from atoms (4), to quantum dots (5), fluorescent molecules (6–9), carbon nanotubes (10–13), and superconducting qubits(14). The cavities can be either photonic microcavities, which can have very high quality factors ($Q$), or surface plasmon resonators, whose mode volume ($V$) can be in the deep subwavelength regime.

To achieve strong or ultrastrong coupling, emitters are typically placed near or in optical cavities, or optical cavities are fabricated around emitters. In either case, the emitter-cavity system is a hybrid system, in which the emitter and cavity are separate objects. In such hybrid systems, the spatial overlap between the emitters and the cavity is often the key factor that limits the light-matter coupling strength. In this work, we generate crystallized films of carbon nanotubes and show that this material exhibits intrinsically ultrastrong interactions. Instead of coupling to external cavities, nanotube excitons (15) couple internally to nanotube plasmon resonances. The nanotubes thus play a dual role as both plasmonic cavities and emitters.

The plasmon resonances considered here comprise longitudinal charge oscillations along the nanotubes coupled to electromagnetic fields (16–23). They are notable for their electrostatic tunability and ability to effectively confine light to a small $V$. As a result of these small volumes, the strong exciton strength of carbon nanotubes, the extremely high density of nanotubes in the crystallized films, and the nearly ideal plasmon-exciton spatial overlap deriving from the intrinsic nature of the cavity-matter interactions, the normalized plasmon-exciton interaction strengths reach $\Omega/\omega_0 = 75\%$. These plasmon-exciton polaritons ("plexcitons") are thus far into the ultrastrong regime, which is typically defined by $\Omega/\omega_0 \gtrsim 30\%$ (2), and one of the most strongly coupled systems that has been achieved in any material.

Crystallized nanotube films could play an exciting role in active, nonlinear optical devices. For instance, they could lead to nanolasers in which the gain medium is intrinsically built into the lasing cavity. Moreover, $\chi^{(3)}$ optical nonlinearities, already strong in dispersed carbon nanotubes, should be dramatically enhanced by ultrastrong plasmon-exciton coupling (24). Many of the technological goals of ultrastrong coupling, including low-power optical modulators and resonant photodetectors, could be more simply achieved with a single material rather than a binary system.

From a fundamental electromagnetics standpoint, the highly anisotropic optical properties of crystallized nanotube films could make them a resource for hyperbolic metamaterials or plasmonic hypercrystals (25) that support low-loss "hyperplasmons." The intrinsic nature of our system's ultrastrong coupling also suggests a natural scalability to the single cavity – single emitter quantum regime. Outside of nanophotonics, the crystallized nanotube films that we fabricate could have diverse applications like high-ampacity conductors (26) and battery anodes. Their successful assembly is a milestone in the larger endeavor of assembling nanostructures into macroscopic functional materials.

**Assembly of crystallized films of carbon nanotubes**

Our crystallized nanotube films are fabricated by adapting and modifying a carbon-nanotube vacuum filtration technique (27, 28). A powder of arc-discharge single-wall carbon nanotubes, of 1.41 nm average diameter, is dispersed in water with the surfactant sodium dodecylbenzenesulfonate (see Methods). Atomic force microscopy confirms that the nanotubes are unbundled in solution (see *SI Appendix*). A very weak vacuum then slowly pulls the suspension through a track-etched polycarbonate membrane, causing the nanotubes to self-organize and align in the plane of the membrane. The membrane is then dried, and the 1 inch-in-diameter nanotube films are transferred to silicon or sapphire substrates.

Surprisingly, the nanotubes in these solution-assembled films can not only be aligned into monolithic, wafer-scalable films, as has been previously shown (27), but can be crystallized into polycrystalline films with large, ~25 nm, domains. Cross-sectional transmission electron microscopy (TEM) of a 200-nm-thick film shows a two-dimensional hexagonal crystal of nanotubes, with at least 50% of the nanotubes crystallized (Figs. 1*A-C* and see *SI Appendix*). Selected area TEM diffraction confirms the hexagonal lattice structure (Fig. 1*D*).

Grazing-incidence X-ray diffraction (XRD) spectroscopy provides further verification of the film's crystallinity (Figs. 1*D* and *SI Appendix*). The prominent XRD peak at $2\theta = 5.8°$, which is 15 times stronger than the XRD peak of a control film of randomly oriented nanotubes, corresponds to an inter-nanotube lattice constant of 1.74 nm and a two-dimensional nanotube density of $3.8 \times 10^5$ μm$^{-2}$. For context, crystallization of nanotubes has also be seen in naturally but randomly produced nanotube ropes generated during nanotube growth (29). However, large-area crystalline films have not previously been observed, nor have crystallized nanotube films been observed to self-assemble in solution.

**Instrinsically ultrastrong plasmon-exciton interactions in crystallized nanotube films**

To investigate their plasmon-exciton interactions (30), we etch the crystallized nanotube films into nanoribbon plasmon resonators (Fig. 2) and study them with μ-Fourier-transform infrared (μ-FTIR) spectroscopy at room temperature. In this geometry, the electric field and charge oscillation of the plasmon modes are polarized along the nanotube alignment direction, with the etched ends of the nanotubes providing confinement. Approximately $10^8$ nanotubes are in each μ-FTIR measurement area (see *SI Appendix*).

The lowest energy semiconducting-nanotube exciton, the $S_{11}$ exciton, has a center energy of $\hbar\omega_0 = 0.66$ eV and an inhomogeneous linewidth of $\hbar\Gamma_0 = 0.2$ eV (Fig. 3*A* and *SI Appendix*). We chemically control the nanotubes' doping level by exposing them to $HNO_3$ vapor, a strong oxidizer, which induces a high density of positive charge carriers. These carriers deplete the nanotube valence band and bleaches the transition strength of $S_{11}$ (31–33). By increasing the free charge density, this surface charge-transfer process also significantly blue-shifts the plasmon resonance energies. We can then reverse this process with vacuum annealing, which eliminates the adsorbates on the nanotubes. After annealing, the on-axis resistivity of our crystallized films increases by a factor of 30 (from $\rho = 2.6 \times 10^{-4}$ ohm·cm to $\rho_0 = 7.8 \times 10^{-3}$ ohm·cm, see *SI Appendix*), and $S_{11}$ is re-strengthened. This resistivity ratio is limited by the presence of metallic nanotubes (~1/3 of the nanotubes) in our films and would be much higher in films of purely semiconducting nanotubes.

In their highly doped state, the nanotube resonators have a strong single extinction peak corresponding to the fundamental transverse magnetic localized surface plasmon resonance (Fig. 3*B*) (20, 22), whose extinction is 99% polarized along the nanotube alignment axis (see *SI Appendix*). As with graphene nanoribbon resonators (34), the resonance frequency ($\omega_p$) is approximately proportional to $\sqrt{qt}$ (22), where the wavevector $q = \pi/L$, $L$ is the nanoribbon width (i.e. the etched length of the nanotubes), and $t$ is the out-of plane thickness of the material (see *SI Appendix*). Since the $S_{11}$ exciton is suppressed in this highly doped state, $\omega_p$ crosses unperturbed through $\omega_0$ as $L$ is reduced.

In more moderate doping states, the plasmon resonance and the $S_{11}$ exciton coexist (see Fig 3*C* and *SI Appendix*). Because the $S_{11}$ excitons are bleached by high free charge densities, and the plasmon resonances require free charges, this coexistence might seem counterintuitive. However, the persistence of the $S_{11}$ exciton in even ambient conditions, where atmospheric adsorbates induce a moderately high doping level, has been observed by many prior studies (15, 32, 33).

In their annealed, resistive state, the nanotube resonators have peaks corresponding to both $\omega_0$ and $\omega_p$. The width of the plasmon resonances, $\hbar\Gamma_p$, ranges from $0.1 - 0.2$ eV and is limited by the etch profile of the nanoribbons. In this annealed state, as *L* is reduced, a large anticrossing between $\omega_p$ and $\omega_0$ becomes evident (Fig. 3*C*). This anticrossing is the key signature of strong coupling. Near the anticrossing, the bare $\omega_p$ and $\omega_0$ resonances hybridize to form $\omega_\pm$ polaritons (Fig. 3*D-E*). The Rabi splitting, $\hbar\Omega$, which we compute as the experimentally measured minimum of $\hbar(\omega_+ - \omega_-)$, reaches 0.485 eV for *t* = 49-nm resonators. In slightly thinner resonators (*t* = 37 nm, see Fig. 4*C*), $\hbar\Omega = 0.50$ eV, and $\Omega/\omega_0 = 75\%$.

The normalized coupling strength between the nanotube plasmon-exciton polaritons is thus significantly beyond the ultrastrong-coupling threshold. In fact, it is among the strongest of any room-temperature system known to date. For context, $\Omega/\omega_0 = 62\%$ has been observed for organic dye molecules are coupled to silver microcavities(8) and $\Omega/\omega_0 = 90\%$ has been observed in III-V planar microcavities (35). For carbon nanotubes coupled to external cavities, $\Omega/\omega_0 = 12\%$ has been attained and $\Omega/\omega_0 \sim 70\%$ predicted to be achievable (12). More recently, $\Omega/\omega_0 = 13.3\%$ has been achieved for carbon nanotubes embedded in planar microcavities (13). At cryogenic temperatures, the record cavity-matter coupling strength is $\Omega/\omega_0 = 130\%$, achieved using superconducting qubits coupled to superconducting resonators (14).

**Tuning the plasmon-exciton interaction strength**

A distinctive feature of carbon-nanotube plasmons is their exceptional tunability. Length, thickness, and doping level are all tuning factors, allowing $\omega_p$ to span frequencies from the terahertz up to the near infrared. Because doping level determines the exciton-transition strength, it modifies not only $\omega_p$ but also $\Omega$ (Fig. 4*A,B*). In turn, this tuning of $\Omega$ allows access to a broad range of polariton energies. This access could translate into electrically tunable optoelectronic devices like photodetectors (36, 37), lasers, and quasi-coherent incandescent light sources (38).

Due to coupled antenna effects, increasing the thickness of the nanotube film leads to higher energy plasmon resonances (22). The plasmon-exciton anticrossing can therefore reached at smaller wavevectors (i.e. higher *L* values) (Fig. 4*C*). Since the plasmonic mode volumes of thicker films are larger than those of thinner ones, their Rabi splittings are also modestly lower, though still extremely strong: a 260 nm-thick film exhibits $\Omega/\omega_0 = 60\%$.

Nonetheless, these thick films should be particularly suitable for optoelectronics due to their large dipole strength. For very thin films (e.g. the $t = 15$ nm film in Fig. 4*C*), $\omega_p$ is much less than $\omega_0$ at even the shortest nanotube length that we could etch ($L = 80$ nm). In this case, due to the weak plasmon-exciton coupling, the higher energy resonance is nearly purely excitonic and barely shifts with *L* (see *SI Appendix*). This behavior is consistent with the fact that the exciton energies of isolated nanotubes do not observably shift with *L* (39).

The plasmon-exciton coupling can be described by a two-coupled-oscillator Hamiltonian:

$$H = \omega_p(a^\dagger a) + \omega_0(b^\dagger b) + \Omega(a^\dagger + a)(b^\dagger + b) + \Omega^2/\omega_0(a^\dagger + a)^2$$

[Eq. 1]

where $a^\dagger$ ($a$) and $b^\dagger$ ($b$) are the exciton and plasmon creation (annihilation) operators, respectively (40, 41). The first two terms represent exciton and plasmon self-energies, respectively, and the second two terms represent plasmon-exciton interactions. Equation 1 includes counter-rotating interaction terms, which are neglected in the frequently used Jaynes-Cummings Hamiltonian, an approximation to Eq. 1, but must be considered in ultrastrong coupling. Neglecting the contribution from linewidths to the polariton energies (see Methods), the full Hopfield solution (9, 40, 41) to Eq. 1 is:

$$\omega^4 - \omega^2(\omega_p^2 + \omega_0^2 + g^2) + \omega_p^2 \omega_0^2 = 0$$

[Eq. 2]

where $g$ is the plasmon-exciton interaction strength, and we have neglected imaginary terms associated with the finite linewidths(9). To fit our data, we use Eq. 2, fix $\omega_0$ to its experimentally observed value of 0.66 eV, and parametrize $\omega_p$ as $X_0/\sqrt{L}$ and $g$ as $X_1/\sqrt{L}$, where $X_0$ and $X_1$ are fitting factors (see Methods). Although this model is certainly only phenomenological, it exhibits excellent agreement with all of our experimental data (see Figs. 3*D*, 4*A*, and 4*C*).

Highly conductive nanotube films can also exhibit strong light-matter interactions. At high energies, the $\omega_p \propto 1/\sqrt{L}$ relationship breaks down, and $\hbar\omega_p$ saturates at 0.9 eV (Fig. 4*D*). Though $S_{11}$ is nearly completely suppressed at high doping levels, many higher energy nanotube excitons are not (see *SI Appendix*). Consequently, even at high nanotube doping levels, the plasmons strongly interact with these higher energy excitons and have flat, saturating dispersion relationships. In the future, smaller diameter nanotubes, which have higher energy optical transitions (10), could allow the 0.9 eV threshold to be crossed. However, our currently achieved tuning range already allows carbon-nanotube plasmonics to be applied to C-band near-infrared telecommunications.

**Conclusions**

Intrinsically ultrastrong coupling represents a compelling new concept in active nanophotonics. It could drive the next generation of tunable infrared devices, including nanoscale light sources, multispectral detectors, and wavefront-shaping chips based on tunable metamaterials (24) and metasurfaces. As ultrastrongly coupled systems, nanotube antennas could serve as physical representations of the Dicke Hamiltonian and a testbed for its quantum-phase transition to superradiance (42). In turn, superradiating antennas could function as bright single-photon sources for quantum communications. Crystals of nanotubes could thus serve as a bridge between fundamental concepts in quantum optics and practical technologies.

**Methods**

**Fabrication of crystallized carbon-nanotube films**

A vacuum filtration method is used to prepare the crystalline carbon nanotube films from nanotube solutions. This method is a modified version of the method used in Refs. (22) and (27). The setup consists of a fritted glass filter and a 15 mL funnel (Millipore XX1002500) purchased from Fisher Scientific Company. The vacuum filtration method consists of four steps:

1. Dispersing nanotubes in a surfactant solution
2. Modifying the surface of a filter membrane
3. Vacuum-filtering the nanotube suspension through the modified membrane
4. Transferring the nanotube film onto a target substrate

An arc-discharge powder of single wall carbon nanotubes, with average diameter of 1.41 nm, 2:1 semiconducting:metallic fraction, and an average length of 500 nm (P2-SWCNT, Carbon Solutions, Inc.), is dispersed into an aqueous solution of 0.4% (wt/vol) sodium dodecylbenesulfonate (SDBS, Sigma-Aldrich). The initial nanotube concentration is 0.4 mg/ml. Bath sonication is applied to the suspension for 15 minutes, followed by tip sonication for 45 minutes. During tip sonication, the suspension is immersed in cold water to prevent heating. The suspension is then centrifuged for 1 hour at 38,000 r.p.m (178,000 × $g$) to remove any remaining nanotube bundles and amorphous carbon. The supernatant is then diluted by a factor of 5 with deionized water, making the SDBS concentration approximately 0.08%. The suspension is then further diluted by a factor of 3 with 0.08% SDBS solution.

The filtration membranes for the vacuum filtration process (Whatman Nuclepore track-etched polycarbonate hydrophilic membranes, 0.05 µm pore size) are first treated with a 2 Torr air plasma for 30 seconds. This treatment provides a negative charge surface on the membrane, which proved to be an important step for achieving a high degree of nanotube alignment and high packing density. Oxygen plasma was also used, but whereas oxygen plasma would etch the membrane, air plasma was gentler and yielded better results.

The nanotube suspension is then filtered through the plasma-treated membranes. For the first 3 mL solution, a weak vacuum pressure (2.8 Torr) is used to pull the suspension through the membrane at a very slow rate, approximately 0.4 mL/hour. This slow filtration speed gives the nanotubes enough time to align and to cover the whole membrane uniformly. After 3 mL has been pulled through the membrane, the vacuum pressure is increased to 8.4 Torr, which results in a filtration rate of 0.8 mL/hour. Finally, before the meniscus contacts the membrane, a high vacuum pressure of 370 Torr is applied to dry the liquid. The film thickness is observed to be proportional to the volume of the precursor used (14 nm per 1 mL of suspension). *SI Appendix* Fig. S2 shows a photo of a membrane with a nanotube film.

The substrates used in this work are high-resistivity silicon and sapphire wafers. For the silicon substrates, the native oxide is first removed by buffered hydrofluoric acid to reduce the coupling of nanotube plasmons to optical phonons in the oxide (20). A drop of water is placed on the substrate, and the membrane is then placed on the substrate with the nanotube surface face down. The membrane is then covered by a glass slide, and gentle pressure is applied to make the nanotube film adhere. The membrane is then dried with gentle $N_2$ gas and then dissolved in chloroform. To remove the surfactant and polymer residue, the nanotube film on the substrate is then annealed at 500 °C in a vacuum oven at $10^{-7}$ Torr for 2 hours.

**Fabrication of nanoribbons and charge-transfer doping**

Conventional electron-beam lithography is used to pattern the CNT films. A bilayer resist consisting of a poly(methyl methacrylate) (PMMA) layer and a hydrogen silsesquioxane (HSQ) layer was spin-coated on the nanotube film. After e-beam exposure, the HSQ is developed as a hard mask. The PMMA layer and nanotubes are then etched with oxygen reactive ion etching. The residual PMMA and HSQ are then stripped with chloroform. The strip length to spacing ratio is fixed at 1:1, as shown in Fig. 2*A*.

To induce strong *p*-type surface charge transfer doping, the sample is exposed to $HNO_3$ vapor for 10 minutes. For a moderate doping level, the sample is first strongly doped and then annealed on a hot plate at 120 °C. The doping conditions 1-3 in Fig. 4*A* are realized with respective heating times of 2, 7, and 30 minutes.

**Characterization of crystallized nanotube films**

The crystalline properties of nanotube films are characterized with a Siemens D5000 XRD spectrometer. The conductivities are measured by Keithley 2400 equipped with an in-line 4-probe system. The extinction spectra are all measured in a Bruker Nicolet 8700 μ-FTIR system, except for the high frequency plasmon resonances in thick films ($\hbar\omega_p > 0.85$ eV, see Fig. 4*D*), which are measured in PerkinElmer Lambda 950 UV-VIS-NIR spectrometer.

**Fitting the data to the Hopfield model**

To perform the fits to Eq. 2, we fix $\omega_0$ to its experimentally observed value of 0.66 eV/$\hbar$. We then parameterize $\omega_p$ as $X_0/\sqrt{L}$ and $g$ as $X_0/\sqrt{L}$, where $X_0$ and $X_1$ are fitting parameters. The $\omega_p$ parameterization derives from the dispersion of a quasi-2D nanoribbon (3, 34).

The parameterization for $g$ follows from the following reasoning. For strongly coupled ensembles of emitters, the maximum cavity-emitter coupling strength is $g \propto \sqrt{Nf/\varepsilon V}$, where $N$ is the number of emitters, $f$ is their oscillator strength, $\varepsilon$ is the dielectric constant, and $V$ is the mode volume (40). In our system, $N$ is the number of nanotubes in a resonator and independent of $L$. We also approximate $f$ and $\varepsilon$ as constants with respect to $L$, and we approximate the plasmon-exciton overlap as a constant.

To paramaterize $V$, we model our plasmon resonators as quasi-2D resonators and apply the same treatment that is typically given to graphene nanoribbon plasmon resonators (3, 34). As fundamental modes, the plasmon modes we are considering extend a distance of $\lambda_{sp}$, the plasmon wavelength, in the nanotube alignment direction. Their extent in the out-of plane direction is also

~ $\lambda_{sp}$ (3, 34). In the in-plane but perpendicular to nanotube alignment direction, the nanoribbons are quasi-infinite, making the extent of the plasmon mode constant in this dimension with respect to $L$. Thus, as for graphene nanoribbons, $V \sim \lambda_{sp}^2$. Using the 2-D dispersion relationship, $\lambda_{sp} \sim \sqrt{L}$, and $g \propto \sqrt{Nf/\varepsilon V}$, we then arrive at $g \propto 1/\sqrt{L}$.

The full Hopfield solution to Eq. 1 includes not only the terms in Eq. 2, but also imaginary terms representing the linewidth of the oscillators. These terms can reduce the plasmon-exciton coupling strength. However, because the plasmon and exciton linewidth are both significantly smaller than $\omega_0$, $\omega_p$ and $g$, we neglect them in Eq. 2.

With $X_0$ and $X_1$ as fitting parameters and $\omega_0$ a fixed, experimentally derived parameter, we then fit Eq. 2 to the experimental data. To perform these fits, we used Matlab, by Mathworks, Inc., and its nonlinear least squares fitting algorithm.

**Acknowledgments**: The authors thank Kuan-Chang Chiu, Jerry Tersoff, James Hannon, Jessie Rosenberg, Weilu Gao, and Junichiro Kono for helpful discussions. This work was funded by IBM and the Postdoctoral Research Abroad Program of the Ministry of Science and Technology Taiwan (National Science Council 106-2917-I-564-012).

**Authors contributions**: P.H. fabricated the crystallized films. P.H. and A.F. performed the optical measurements. P.H. and G.T. purified the nanotube suspensions. L.M.G. performed the TEM measurements. J.B. performed the electron-beam lithography. P.H. and D.M.B. performed the XRD measurements. All authors participated in analyzing the data and writing the paper.

This article contains supporting information online.

The authors declare no competing interests.

**Figures**

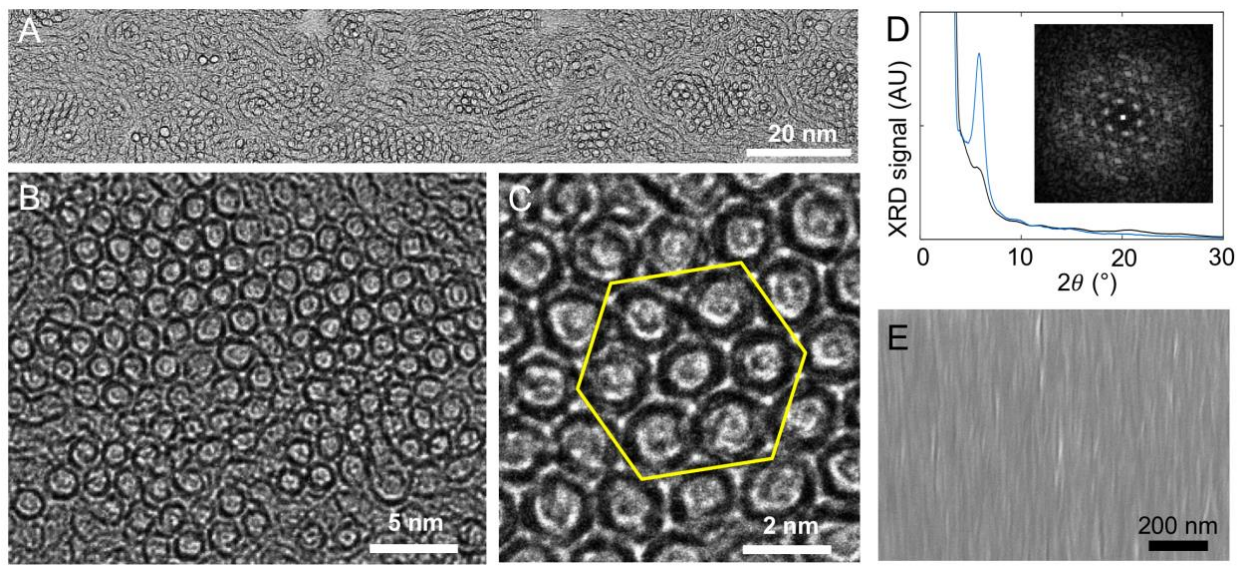

**Fig. 1.** Crystallized carbon-nanotube films. (*A-C*) Cross-sectional TEM images of the crystallized nanotube films at three magnifications. (*D*) Grazing-incidence XRD spectra of a crystallized nanotube film (blue curve) and a control film of randomly oriented nanotubes (black curve). The peak at $2\theta = 5.8°$ corresponds to an inter-nanotube lattice spacing of 1.74 nm. The x-ray beam size is 2 cm$^2$. Inset: Selected-area electron diffraction image of the region in Fig. 1*C* confirming the hexagonal lattice structure. (*E*) Scanning electron micrograph of the top surface of a crystallized nanotube film. The nanotube alignment axis is vertical.

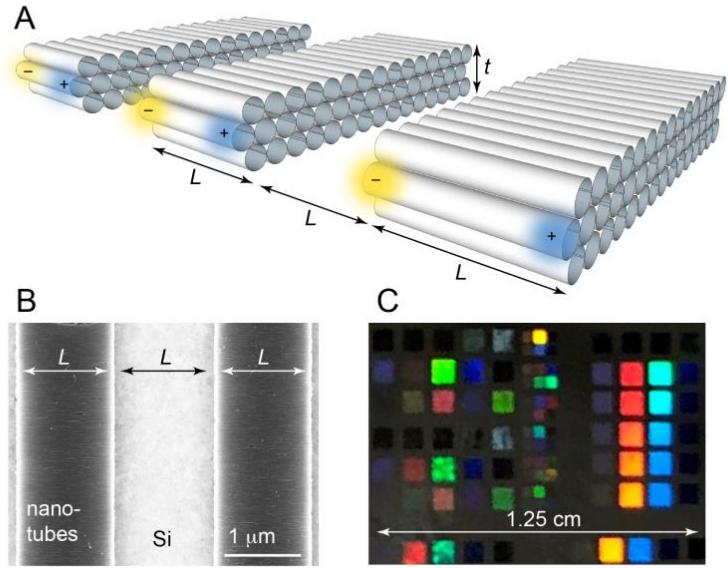

**Fig. 2.** (*A*) Plasmonic resonators comprising crystallized nanotube films etched into nanotube nanoribbons. The etched nanotube length is *L* and inter-ribbon spacing is *L*. For the depicted transverse-magnetic plasmon resonance, the electric field and charge oscillation are both parallel to the nanotube-alignment axis. (*B*) A scanning electron micrograph of a crystallized nanotube film that has been etched into plasmon resonators. The nanotube alignment axis is horizontal. (*C*) A photograph of a crystallized film after it is etched. The color of diffracted light varies with the pitch (2*L*, see Figs. 2*A*, *B*) of the grating. Each 1 mm or 0.5 mm square in this photograph consists of a one-dimensional array of etched nanotube nanoribbons, with *L* ranging from 80 nm to 2 μm, and the pitch of each array being 2*L*.

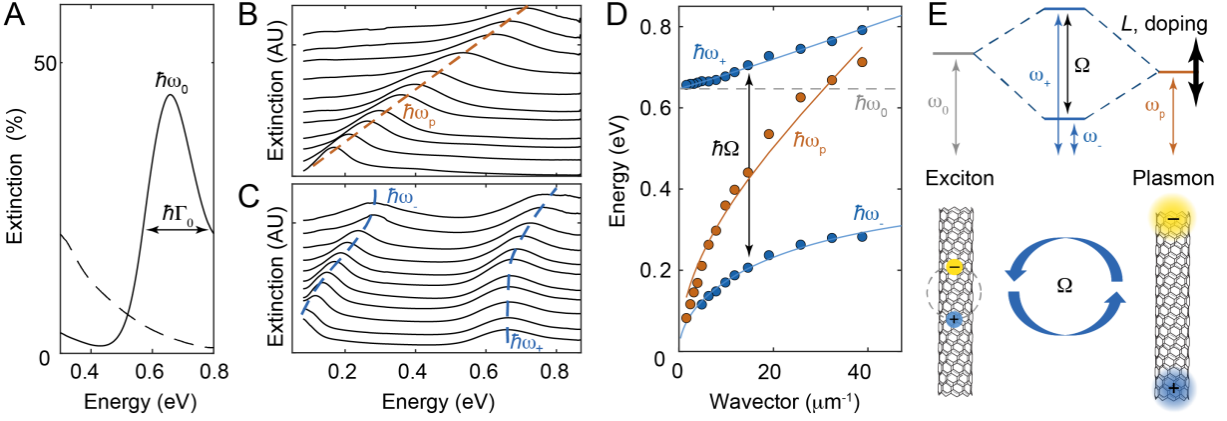

**Fig. 3.** Intrinsically ultrastrong plasmon-exciton interactions. (*A*) The extinction spectra of a crystallized nanotube film in its annealed, resistive state (solid line), showing a strong $S_{11}$-exciton peak, and a highly chemically doped film (dashed line), where $S_{11}$ is bleached. The film thickness is $t = 49$ nm. (*B*) The extinction spectra of plasmon resonators etched into this film, with a fixed, high doping level and varying lithographically defined nanotube lengths ($L$ values). For clarity, each curve is vertically offset and normalized by its maximum. (See *SI Appendix* Fig. S5 for un-normalized curves). From top to bottom, $L = $ (80, 100, 130, 175, 225, 260, 325, 400, 500, 650, 800) nm. The dashed line is a guide for the eye. (*C*) In their annealed, resistive state, the nanotube resonators have extinction spectra with two peaks each, corresponding to excitons and plasmons that hybridize to form polaritons. (*D*) Orange: Peak energy of the highly doped plasmon resonators from Fig. 3*B* vs. wavevector ($q$), defined as $q = \pi/L$. The solid line is a fit to $\omega_p \propto \sqrt{q}$. Blue: Peak energies of the plasmon-exciton polaritons from Fig. 3*C*, showing an anticrossing with a Rabi splitting of $\Omega = 0.485$ eV. The solid line is a fit to Eq. 2. (*E*) Energy-level diagram and depiction of the nanotube excitons hybridizing with the nanotube plasmons to form ultrastrongly coupled plasmon-exciton polaritons.

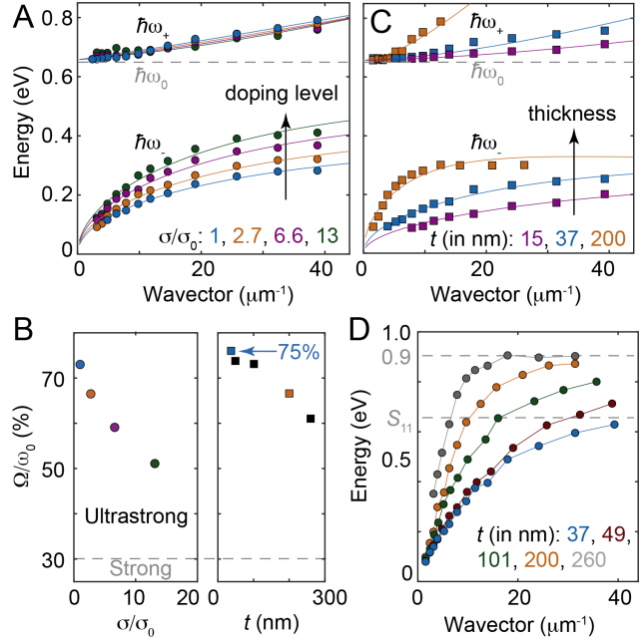

**Fig. 4.** Tunability of plasmon-exciton interactions. (*A*) Polariton energies vs. wavevector for resonators at 4 different doping levels. The film conductivity for each curve is specified in the legend as a multiple of $\sigma_0 = 1/\rho_0$. The film thickness is $t = 49$ nm. (*B*) Left: $\Omega$ vs. film conductivity ($\sigma$), with $t$ fixed to 49 nm. Right: $\Omega$ vs $t$, with $\sigma$ fixed to $\sigma_0$ (i.e. immediately after the film has been vacuum annealed). The colored points are color-coded to match the curves in Figs 4*A* and 4*C*. (*C*) Polariton energies of resonators as a function of $q$ and $t$, with the doping level at its minimum ($\sigma = \sigma_0$). (*D*) Plasmon resonator energies when the nanotubes are maximally doped (i.e. immediately after $HNO_3$ exposure). When $q$ and the doping level are both high, $\omega_P$ saturates at 0.9 eV.

# Supplementary Information for

## Intrinsically Ultrastrong Plasmon-Exciton Interactions in Crystalline Films of Carbon Nanotubes

Po-Hsun Ho, Damon B. Farmer, George S. Tulevski, Shu-Jen Han, Douglas M. Bishop, Lynne M. Gignac, Jim Bucchignano, Phaedon Avouris, Abram L. Falk

Abram L. Falk  
Email: alfalk@us.ibm.com

Phaedon Avouris  
Email: phaedon.avouris@gmail.com

**This PDF file includes:**

Supplementary text

    S1. Properties of the nanotube suspension

    S2. Crystal structure of the nanotube films

    S3. Spectroscopy methods

    S4. Optical transitions of the crystalline nanotube films

    S5. Resistivity of the crystalline nanotube films

    S6. Quasi-2D dispersion of the nanotube nanoribbons

    S7. Linear dichroism of the nanotube resonators

    S8. Variation of the spacing between nanotube plasmon resonators

    S9. Association between upper and lower plasmon-exciton polariton branches

Figs. S1 to S12

## S1. Properties of the nanotube suspension

To verify that the nanotubes are indeed well dispersed in SDBS suspension, we have checked this suspension by drop-casting it on blank silicon chips, drying the chips, and imaging them with atomic force microscopy. No nanotubes whose height is greater than 2 nm are observed (Fig. S1), indicating that our sonication and centrifugation processes are effective at removing bundles from the suspension.

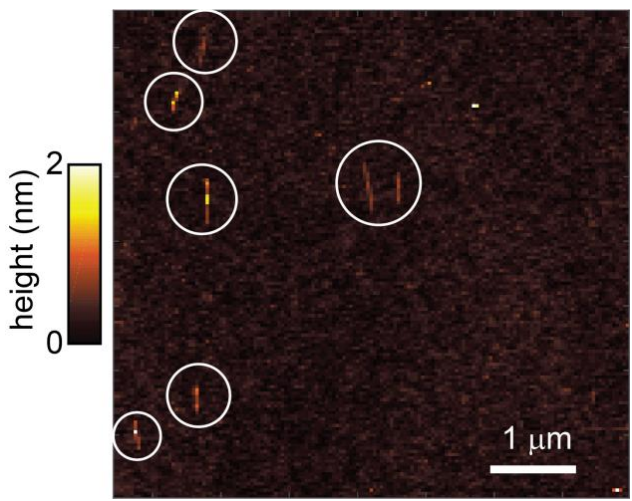

**Fig. S1.** Atomic force microscopy (AFM) image of nanotubes that are drop-casted and dried on a silicon chip. They are from an aqueous SDBS suspension that was used to fabricate our crystallized nanotube films.

## S2. Crystal structure of the nanotube films

With the modified vacuum filtration method, a one-inch crystalline nanotube film was deposited on a polycarbonate membrane (Fig. S2). As discussed in SI Ref. (27), this process can potentially be readily extend to the wafer scale by using a larger fritted glass and membrane. Both the microscopic and macroscopic crystalline properties of this nantoube film were measured by selective area electron diffraction and grazing-incidence X-ray diffraction.

Fig. S3*A* shows a cross-sectional TEM image of a crystalline nanotube film. We measured the selective area diffraction pattern in regions of this image (Fig. S3*A*). The corresponding diffraction patterns are shown in Fig. S3*B*. Different crystal angles of hexagonal lattice were observed, indicating that this nanotube film is polycrystalline.

Grazing incidence X-ray diffraction (XRD) was used to further characterize the crystallinity of the aligned CNT arrays. The reason that the grazing incidence mode was chosen is that it is particularly effective at measuring thin films. Here, we focus on the low angle diffraction related to the >1 nm lattice constant from the lattice of nanotube (as opposed to the much higher angle diffraction coming from the carbon lattices of the comprising nanotubes).

The XRD was performed with a Siemens D5000 XRD spectrometer and a Cu-Kα ($\lambda$ = 1.54 Å) x-ray source. The incident angle was fixed at 2.2°, and the angle of the detector was swept. A strong discrete peak is observed at 2 = 5.8°, which corresponds with the lattice spacing of the aligned nanotube arrays. Based on the two-dimensional hexagonal lattice of nanotubes indicated by the selected area TEM diffraction (Fig. 1*C*, 1*D*, and Fig. S3*B*), this signal comes from the (10) plane of the lattice. The $d_{(10)}/a$ ratio is $\sqrt{3/2}$, where $a$ is the lattice constant. Bragg's law, $2d \sin\theta = n\lambda$, therefore indicates that the lattice constant of the nanotube lattice is $a$ = 1.74 nm. In light of the $D$ = 1.41 nm average diameter of nanotubes, as specified by the nanotube manufacturer, the van der Waals gap between neighboring nanotubes, $a - D$, is 3.3 Å.

XRD measurements with the nanotube sample oriented in different directions relative to the incident light were also performed (Fig. S4*A*). The collected data are all indexed and shown in Fig. S4*B*. When the CNTs are parallel to the incident plane, the strongest (10) and (11) signals were obtained. However, the higher index signals were smaller than those corresponding to other alignment directions. These results confirm the preferred orientation direction of the polycrystalline CNT film.

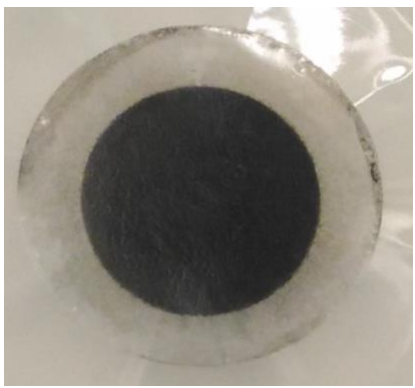

**Fig. S2.** A photograph of a 1-inch-in-diameter CNT film on a membrane.

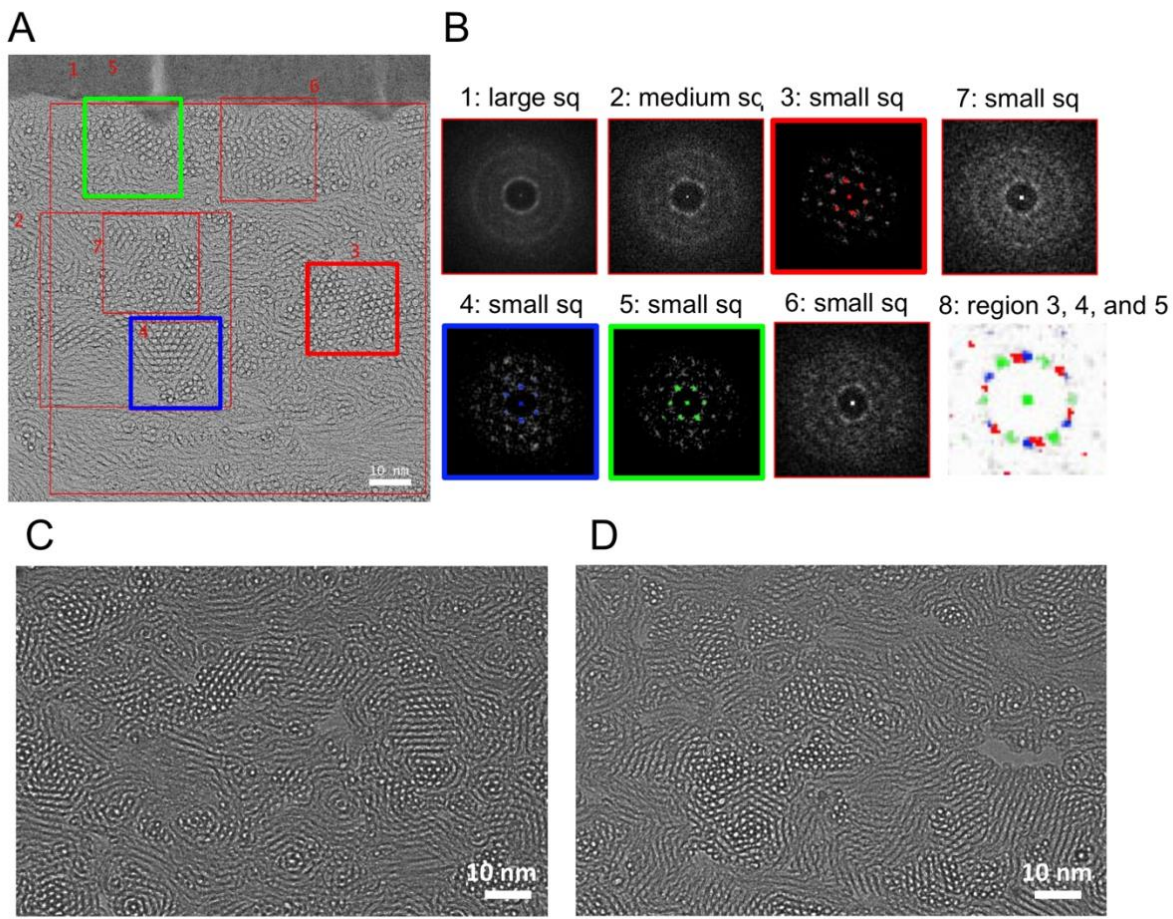

**Fig. S3.** (*A*) A cross-sectional TEM image of the nanotube film shown in Figs. 1*A*-*C*. (*B*) Selective area electron diffraction patterns (SAED), corresponding to the regions marked in Fig. S3*A*. In image #8, the locations of the diffraction spots regions 3, 4, and 5 are combined. (*C*) and (*D*) Cross-sectional TEM images of another nanotube film, also showing clear polycrystallinity.

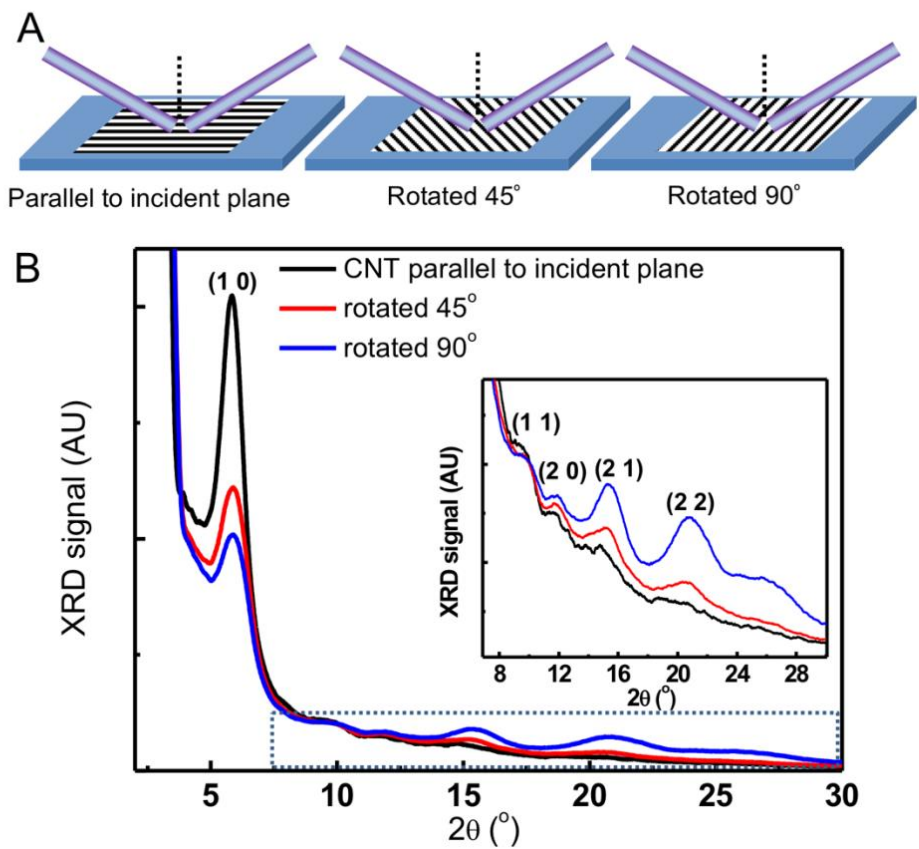

**Fig. S4**. Angle-dependent XRD measurements. (*A*) Schematic images of XRD measurements with different orientations of the CNT film (the black lines) relative to the x-ray beam. (*B*) XRD profile of the CNT film with different orientations. Inset: magnified image of the high angle region (dashed box in the main figure panel).

## S3. Spectroscopy methods

The extinction spectra were measured in a Bruker Nicolet 8700 µ-FTIR system. The diameter of the focused beam size is 100 µm, and each measurement was integrated for a 100-second period. Before each spectrum is taken, a background spectrum is taken, on an area of the chip in which the carbon nanotubes have been completely etched away. The spectrometer outputs absorbance relative to the background spectrum. The extinction is then calculated as

$$Extinction = 1 - 10^{-absorbance(v)}. \hspace{2cm} [S1]$$

Measurements of the high frequency plasmon resonances in thick films ($\hbar\omega_P > 0.85$ eV, see Fig. 4*A*), which were out of the range of our FTIR system, were executed using a PerkinElmer Lambda 950 UV-VIS-NIR spectrometer. The films we prepared for use in this system were similar to the ones that we prepared for FTIR, except 1) larger patterns (3 mm × 7 mm) were prepared, to accommodate the larger beam size of this spectrometer, and 2) sapphire substrates were used instead of silicon, to extend the range to higher energies.

Fig. S5 presents the non-normalized (i.e. as-measured) extinction spectra corresponding to Figs. 2*C* and 2*D*. In both spectra, we observe that as the lithographically defined CNT length (*L*) becomes shorter, the extinction signal becomes less, even though a 1:1 nanotube:gap coverage ratio is maintained. The simplest explanation for this phenomenon is that, as *L* is reduced, there is an increasing wavevector mismatch between the nanotube resonators and free space. Another contributing factor may be damaged ends of the nanotubes, which arise from the reactive ion etching, reducing the absorption of the shorter-*L* nanotubes.

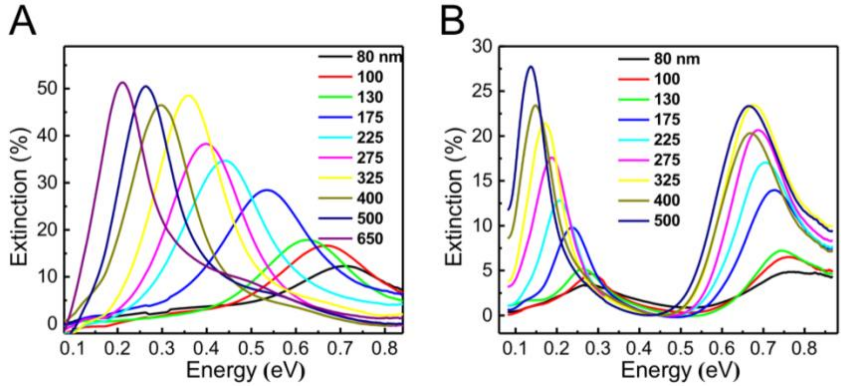

**Fig. S5.** Extinction spectra without normalization. (*A*) The nanotube resonators are in a high doping-level state. The film thickness is *t* = 49 nm. The etched nanotube length (*L*) is given in the legend. (*B*) The nanotube resonators are in a low doping-level (annealed) state. In Figs. 3*B-C* of the main text, we normalized the height of the extinction spectra and offset them from each other, for clarity of presentation. Here, we present the same data, except that we do not normalize the curves or offset them from one another.

## S4. Optical transitions of the crystalline nanotube films

We used an FTIR spectrometer to perform all of the extinction spectroscopy measurements in the main text. However, this instrument only measures energies up to 0.95 eV. To measure higher energy optical transitions in our crystalline nanotube films, we used a PerkinElmer Lambda 950 UV-VIS-NIR spectrometer. For these measurements, to avoid silicon absorption, we used sapphire substrates. The detection area of this system is 25 mm$^2$.

Figure S6 presents the UV-VIS spectra of a 20 nm-thick film with different doping levels. Three excitons can be clearly identified: the $S_{11}$ and $S_{22}$ excitons of semiconducting CNTs, and the $M_{11}$ exciton of metallic CNTs. With increasing $p$-type free charge density, the $S_{11}$ excitons are quenched first, due to depopulation of the highest-energy valence states. The $S_{22}$ exciton is next to be quenched. At very high doping levels, the selection rules of the optical transitions are modified, allowing new inter-valance-band transitions (32). Therefore, new optical transitions can be observed. In our CNT film, we observe an inter-valence band transition $S_{41}$, whose energy is 1.04 eV. This state has previously been observed in strongly doped CNT films (32, 33). These inter-valance excitons persist at high $p$-type doping levels and can strongly couple to plasmon resonances. They are likely responsible for the saturating resonance energies at 0.9 eV that are observed in Fig. 4$D$ of the main text.

The strength of the interband excitons and the plasmon resonances have opposed doping dependences. Because plasmons comprise charge oscillations, the plasmon resonance strengthens as the doping level increases and more free charges are available. On the other hand, an increase in the doping level causes a depopulation of the band-edge states that leads to a smaller density of available transitions for interband nanotube excitons. Thus, the exciton transition strength decreases with increasing doping level.

Nonetheless, there is a substantial regime of moderate doping levels in which the exciton and the plasmon resonance are both strong. As Fig. S6 and S7 show, while the $S_{11}$ exciton is eventually quenched at high doping levels, it persists and is reasonably strong at a variety of moderate doping levels. This persistence is confirmed by absorption measurements in previous studies (32, 33), as well as two-photon fluorescence studies of nanotube exciton binding energies (15) that are performed in ambient conditions, where nanotubes are doped by surface adsorbates.

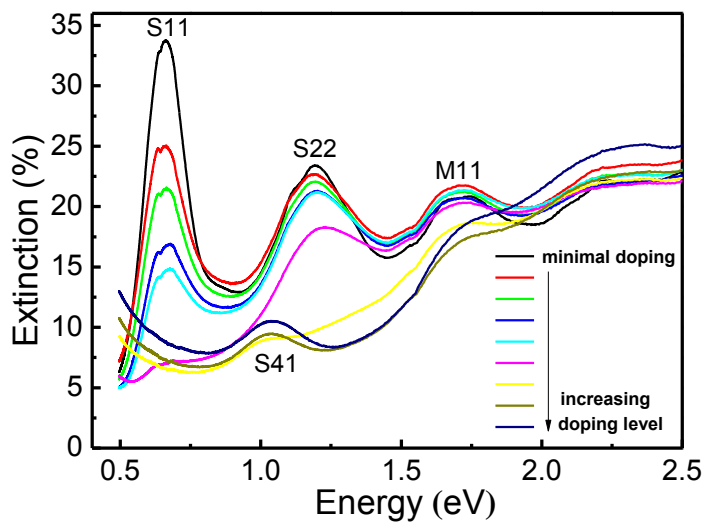

**Fig. S6**. UV-VIS extinction spectra of an unpatterned 20-nm-thick crystallized nanotube film, as a function of chemically induced charge-carrier density. Although $S_{11}$ and $S_{22}$ decrease in intensity with increasing doping level, a new peak, $S_{41}$, emerges at very high doping levels (32, 33).

As for the plasmon resonance, although it strongly redshifts with lower doping levels, it remains reasonably strong at all doping levels that we can access, even in the "annealed, resistive" state of our patterned films, which is the state of our films in Figs. 3*C-D* and Fig. 4*C*. We note that the conductivity in this state is only 30 times lower than that of the most strongly doped state – a much lower on/off ratio than that of a pure semiconducting nanotube or a film of purely semiconducting nanotubes.

Figure S7 shows the coexistence of the plasmon and exciton resonances in a patterned crystallized nanotube film. In this film, $t = 49$ nm and $L = 650$ nm. The long *L* value is chosen so that $\omega_p \ll \omega_0$, and thus, the plasmons and excitons are not hybridized.

For the highest doping level (the turquoise curve), which is the state of the film immediately after $HNO_3$ vapor exposure, the $S_{11}$ exciton transition strength at $\omega_0$ is quenched. Here, there is only a single extinction peak, which corresponds to the plasmon resonance. However, at the four lower doping levels, the exciton and plasmon can be seen to coexist. For the three lowest doping levels, extinction from both the exciton and plasmon can be seen to be strong.

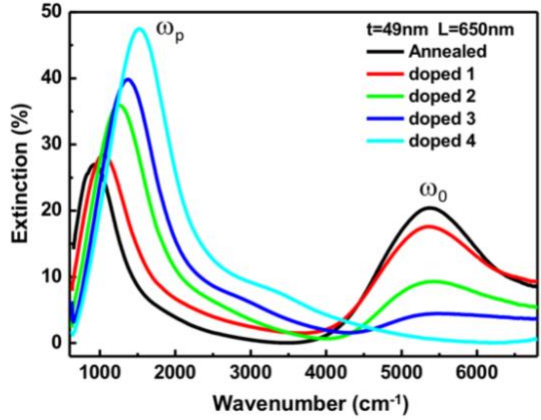

**Fig. S7.** FTIR extinction spectra of nanotube resonators with varying doping densities. The film is patterned into resonators (see Fig. 2) with $L = 650$ nm. With this long $L$-value, the plasmons and excitons are uncoupled. This figure shows that there is a substantial doping regime in which both strong plasmon resonances and strong exciton absorption can both be present.

## S5. Resistivity of the crystalline nanotube films

An in-line four probe method was used to measure the resistivity of our crystallized CNT films. For a film of thickness $t = 100$ nm, in its minimal doping state (i.e. measured right after it has been annealed in vacuum), the as-measured four-probe resistance is $R_x = 171$ Ω, where $R_x$ corresponds to the resistance in the nanotube alignment axis. The four-probe resistance measured perpendicular to the alignment axis ($R_y$) is four times higher than $R_x$. After the nanotube films are exposed to HNO$_3$ vapor, when the nanotube films are in a highly chemically doped state, $R_x$ is reduced by a factor of 30.

To calculate the sheet resistivity, we model the CNT film as an infinite 2D sheet, and calculate the resistivity of nanotube film as $\rho_x = \frac{\pi}{\ln(2)} R_x t$. Using this formula, we calculate on-axis sheet resistivities of $7.6 \times 10^{-3}$ Ω·cm (after vacuum annealing) and $2.6 \times 10^{-4}$ Ω·cm (after HNO$_3$ vapor treatment). We note that, due to the anisotropy of the film, the factor of $\frac{\pi}{\ln 2}$, which accounts for the spread of current in the sheet, overestimates the true current spread in our film. The resistivities that we report should thus be understood as conservative estimates that modestly overestimate the true resistivity of our films.

## S6. Linear dichroism of the nanotube resonators

Figure S8 shows the exctinction spectra of a film of plasmon resonators as a function of incident light polarization. The peak absorption is 20%, when the incident light polarization is parallel to the nanotube alignment axis, vs. approximately 0.15%, when the light is perpendicular to it. Accordingly, we calculate the degree of linear dichroism of the nanotube resonators to be 99%.

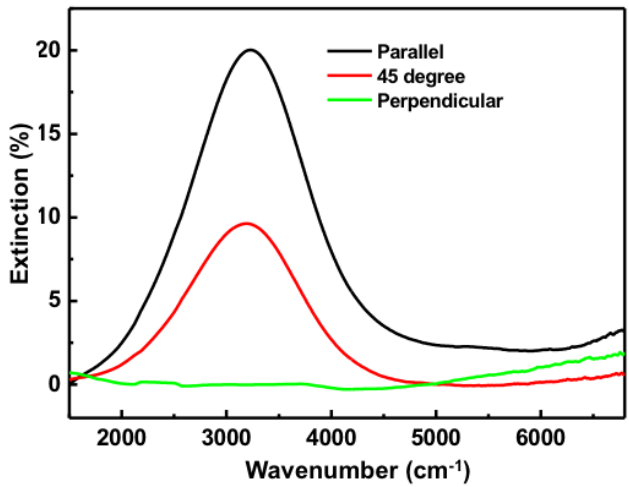

**Fig. S8**. Linear dichroism of the plasmon resonators. Polarization-dependent extinction data of plasmon resonators comprising an etched crystallized nanotube film with $t = 49$ nm and $L = 130$ nm. The nanotubes are in a highly doped state. The polarization of light is varied from being parallel (black curve), 45° (red curve), and perpendicular to the nanotube alignment axis.

## S7. Quasi-2D dispersion of the nanotube nanoribbons

As mentioned in the main text, in the heavily doped state, the plasmon resonance energy of our nanotube nanoribbons follows a $\omega_p \propto \sqrt{tq}$ dependence, where $t$ is the nanotube film thickness (see Fig. 2) and $q = \pi/L$, until $\hbar\omega_p$ saturates at 0.9 eV (see Fig. 4D). We previously experimentally verified this relationship (22), and one can also see the $\sqrt{q}$ relationship in Fig. 3D.

This $\omega_p \propto \sqrt{tq}$ relationship follows from the quasi-2D nature of the nanotube nanoribbons. The $\sqrt{q}$ dependence is also observed in graphene nanoribbons (34), and the $t$-dependence follows from the fact that the sheet-charge density in the nanoribbons is proportional to the film thickness.

Intuitively, the $\sqrt{q}$ dependence can be understood from Fig. S9. In the quasi-2D picture, the plasmon oscillation conists of 1D lines of charges that oscillate parallel to the nanotube alignment axis. From Gauss's law, the attractive electric force between the positive and negative lines of charge, at their maximum separation of *L*, is proportional to 1/*L*. As a simple harmonic oscillator, the resonance frequence of this oscillationnanoribbon has a resonance frequency that is proportional to the square root of this force. Thus, $\omega \propto \sqrt{q}$.

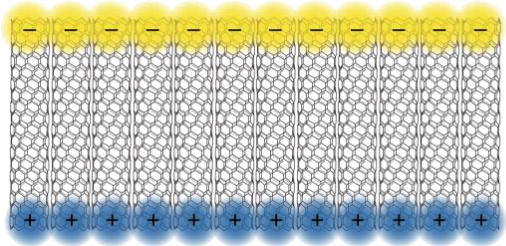

**Fig. S9.** After being etched into nanoribbons, our nanotube films form resonators with a quasi-2D dispersion relationship.

## S8. Variation of the spacing between nanotube plasmon resonators

For all of the data in the main text of this manuscript, the ratio of the etched length of the nanotubes (*L*) and the spacing between them (*s*) is kept fixed at 1:1. Here, in Fig. S10, we show data in which *s* is varied from 130 to 325 nm, while *L* is kept constant at 325 nm. As *s* is decreased, the total absorption in both the exciton and plasmon resonances goes up, a simple consequences of the higher nanotube coverage in the measurement area. However, the center resonance frequency of neither peak changes substantially with *s*. This can be understood as a consequence of the fact that even *s* = 130 nm, the minimum spacing that we could achieve with our oxygen plasma etcher, is a much larger spacing than the inter-nanotube spacing within the crystallized films (1.74 nm). Thus, although our fabricated structures are large gratings of nanotube nanoribbons in order to fill up the measurement area of our FTIR spectrometer, each nanoribbon can be approximately understood as an isolated entity.

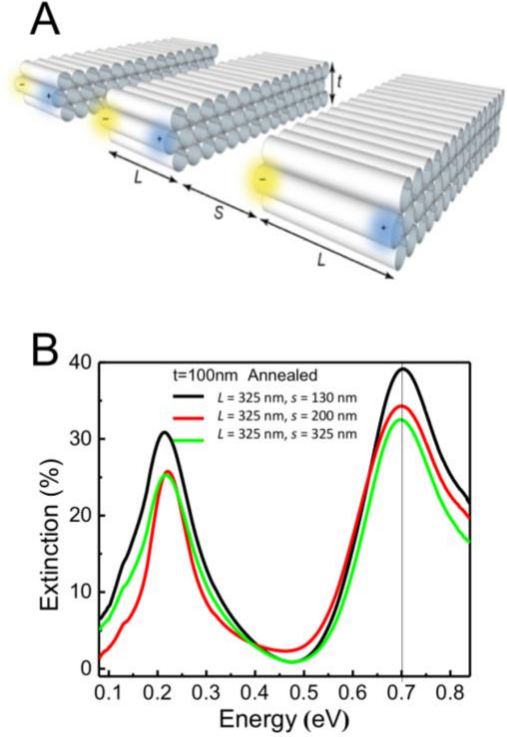

**Fig. S10**. Variation of the spacing between nanotube plasmon resonators. (*A*) Cartoon of the nanotube resonators showing the parameters *L* and *s*. (*B*) Extinction spectra of the nanotube resonators, in a low-doping-level state, with *L* fixed at 325, and *s* varied.

## S9. Association between upper and lower plasmon-exciton polariton branches

The plasmon-exciton Rabi splitting is extremely large in our system. As a result, the fact that the plasmon and exciton branches are truly anticrossing, and not independently evolving with *L*, is important to establish. In this section, we provide a supplementary discussion of the evidence indicating that the apparent anticrossing truly derives from plasmon-exciton coupling, and not from an independent evolution of plasmons and excitons that mimics this coupling.

1. Evolution of $\omega_+$ with *L* is only seen when $\omega_p$ is sufficiently high.

We observe that the frequency of the higher energy peak only deviates from being a constant $\omega_0$ when $\omega_p$ approaches $\omega_0$. In this case, the excitons and plasmons are hybridized, and the lower/higher energy peaks are respectively the $\omega_-/\omega_+$ polaritons. When $\omega_p$ is much less than $\omega_0$, then the excitons and plasmons are not hybridized. In this case, the lower energy peak is just $\omega_p$, and the higher energy peak is $\omega_0$, which stays nearly constant with respect to *q*. For sufficiently small $L$, $\omega_p$ will eventually be high enough such that $\omega_p$ and $\omega_0$ strongly couple. However, for thin films, this $L$ value is much smaller (i.e. the corresponding $q$ value is much higher).

This can be seen in Fig. 4*C* of the main text. Consider the different peak energies of the thinnest film (the $t = 15$ nm film) and the thickest film (the $t = 200$ nm film) at the $q = 13$ μm$^{-1}$ point. For the thin film, the lower energy peak is at 0.12 eV which is 0.17 eV, significantly less than both $\omega_0$ and $\omega_0 - \Omega$. The higher energy peak is at 0.67 eV, which is just 1.5% higher than $\omega_0$ = 0.66 eV. This small degree of departure of the higher energy peak from $\omega_0$ can be understood to result from the fact that the plasmons and excitons are not signficantly hybridized at the $q = 13$ μm$^{-1}$ point for this thin film.

On the other hand, for a thick, $t = 200$ nm film, the excitons and plasmons are significantly hybridized at this $q = 13$ μm$^{-1}$ point. Here, the polariton energies are $\hbar\omega_- = 0.28$ eV and $\hbar\omega_+ =$ 0.79 eV, which is 20% higher than $\omega_0$. The deviation of $\omega_+$ from $\omega_0$ is 10 times stronger than that observed in the thin $t = 15$ nm film at $q = 13$ μm$^{-1}$. This high degree of $\hbar\omega_+$ evolution results from the hybridization of plasmons and excitons.

2. Existing literature has found that, absent cavity coupling, nanotube exciton energies are a constant function of nanotube length.

When plasmons are not coupled to the $S_{11}$ exciton, the $S_{11}$ exciton energy has been observed not to change with $L$ (39). More specifically, in the range of nanotube lengths from 10 nm to 750 nm, Ref S4 finds no discernible change at all in the $S_{11}$ energy.

3. In predominantly metallic nanotube resonators, the exciton energy does not evolve with $L$.

In addition to the films of with mixed-chirality nanotubes (2:1 semiconducting:metallic types) fabricated in the main text, we also fabricated films of aligned carbon nanotubes out of purified metallic nanotubes. We estimate that the residual fraction of semiconducting nanotubes in these films is $S = 2$ to 3%. We purchased these purified nanotubes from Nanointegris, Inc., and fabricated aligned films of metallic nanotubes films in a very similar manner to the films produced in the main text. We then etched these films into gratings and measured the peak energies, much as we did for the films in the main text.

Figure S11 shows the results. As with the $S = 2/3$ films studied in the main text, there is an upper and lower energy peak at each $q$ value. However, for the $S = 2\text{-}3\%$ films, unlike the $S = 2/3$ main-text films films, the upper peak remains constant ($\omega_0$) and does not evolve with $L$. This phenomenon can be understood to derive from the fact that the upper energy peak is nearly a pure exciton, not a plasmon-exciton polariton, because the concentration of semiconducting nanotubes is too low.

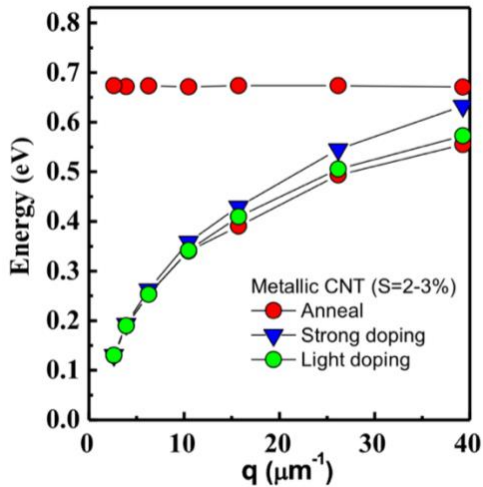

**Fig. S11.** Plasmons and excitons in films of nearly purely metallic nanotubes. The plot shows the two extinction peaks (the $S_{11}$ exciton at 0.67 eV, and the plasmon resonance) of plasmon resonators etched into a $t = 55$ nm film of nanotubes, which are 97-98% metallic and 2-3% semiconducting. The $S_{11}$ exciton, which is only strong enough to be observed when the nanotubes are in their most resistive state, has a constant ($\pm 2\%$) energy with respect to $q$.

4. In thick nanotube films, the lower polariton has a flat shape, characteristic of strong coupling.

As discussed in Supplementary Information section S7, the energy of the nanotube plasmon resonances follows a $\omega_p \propto \sqrt{q}$ relationship. In contrast, for the thick nanotube films, where plasmon-exciton hybridization happens at low $q$ values, the shape of $\omega_-$ clearly deviates from $\sqrt{q}$ and instead, has a flat, saturating shape. For instance, for the $t = 250$ nm film in Fig. 4C, the $\omega_-$ polariton is nearly completely flat by $q = 10$ μm$^{-1}$. This flat shape is a signature of strong coupling.

5. The Rabi splitting can be systematically controlled with doping level and thickness.

As expected from strong plasmon-exciton interactions, thickness and doping level are both parameters that can control $\Omega$ (see Fig. 4B). Naturally, at smaller $\Omega$ values, the fact that $\omega_-$ and $\omega_+$ are anticrossing and not independently evolving becomes more apparent. In Fig. S12, in order to make $\Omega$ as small as we practically can, without quenching the $S_{11}$ exciton with excessive free charge, we use a thick film ($t = 200$ nm) in a moderately doped state.

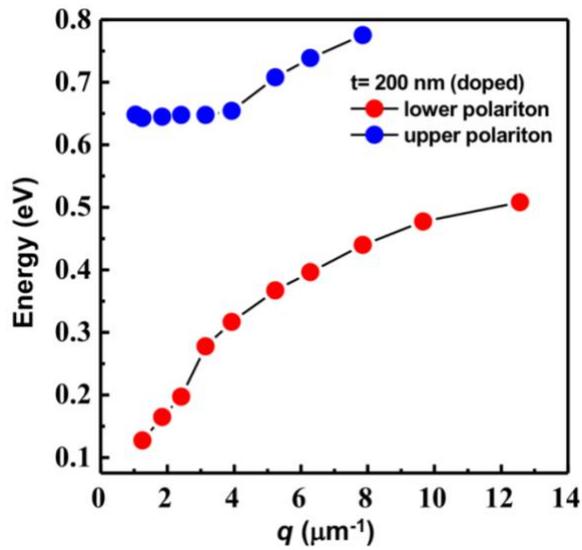

**Fig. S12.** Plasmon-exciton polaritons with a relatively small $\Omega$ in a thick, moderately doped film. To achieve a moderate doping level in this sample, it was exposed to $HNO_3$ vapor and then briefly annealed on a hot plate, so that $\sigma/\sigma_0 = 5$.